# Improved Thin Film Quality and Photoluminescence of N-Doped Epitaxial Germanium-on-Silicon using MOCVD

*Guangnan Zhou[1,*], Alejandra V. Cuervo Covian[2,*], Kwang Hong Lee[3], Chuan Seng Tan[3, 4], Jifeng Liu[2,**] and Guangrui (Maggie) Xia[1,***]*

[1]Department of Materials Engineering, University of British Columbia, 309-6350 Stores Rd, Vancouver, BC V6T1Z4, Canada.

[2]Thayer School of Engineering, Dartmouth College, 14 Engineering Drive, Hanover, New Hampshire 03755, USA.

[3]Low Energy Electronic Systems (LEES), Singapore-MIT Alliance for Research and Technology (SMART), 1 CREATE Way, #10-01 CREATE Tower Singapore 138602.

[4] School of Electrical and Electronic Engineering, Nanyang Technological University, 50 Nanyang Avenue, Singapore 639798.

*Guangnan Zhou and A. V. C. Covian contributed equally to this work.

**Jifeng.Liu@dartmouth.edu

***guangrui.xia@ubc.ca

**ABSTRACT:**

Ge-on-Si structures in-situ doped with phosphorus or arsenic via metal organic chemical vapor deposition (MOCVD) were investigated. Surface roughness, strain, threading dislocation desnity, Si-Ge interdiffusion, dopant diffusion, and photoluminescence were characterized to study the impacts of defect annealing and Si substrate offcut effects on the Ge film quality and most importantly, the light emission properties. All samples have a smooth surface (roughness < 1.5 nm), and the Ge films have a small tensile strain of 0.2%. As-grown P and As-doped Ge films have threading dislocaiton densities from $2.8 \times 10^8$ to $1.1 \times 10^9$ cm$^{-2}$ without defect annealing. With thermal cycling, these values reduced to $1\text{-}1.5 \times 10^8$ cm$^{-2}$. The six degree offcut of the Si substrate was shown to have little impact. In contrast to delta doping, the out-diffusion of dopants has been successfully suppressed to retain the doping concentration upon defect annealing. However, the photoluminescence intensity decreases mostly due to Si-Ge interdiffusion, which also causes a blue-shift in the emission wavelength. Compared to a benckmarking sample from the first Ge laser work doped by delta doping method in 2012, the as-grown P or As-doped Ge films have similar photoluminescence intensity at a 25% doping concentration and smoother surface, which are promising for Ge lasers with better light emission efficiencies.

**Key Words: germanium-on-Si, photoluminescence,**

Germanium (Ge), as the most silicon (Si) compatible semiconductor, are playing an increasingly important role in large-scale dense Si photonic integration, such as in light sensing and modulation [1, 2]. In the past few decades, researchers all over the world have invested extensive efforts in finding solutions to a Si-compatible lasing material system [3-13]. Among all of the candidates, III-V quantum dot (QD) lasers grown on Ge-on-Si substrates and Ge-on-Si lasers [14-16] have been demonstrated to be the most promising on-chip light sources [17].

For Ge-on-Si lasers, important material requirements commonly include 1) very high n-type doping above $1\times10^{19}$ cm$^{-3}$, 2) tensile strain > 0.2% , 3) low threading dislocation density (TDD), 4) surface roughness in nm scale or lower, and 5) minimized Si-Ge interdiffusion at Ge/Si interfaces to prevent detrimental indirect gap behavior. These requirements are also closely related and some can be traded-off such as the doping and strain requirements. Adding high concentrations of n-type dopants in Ge is crucial to occupy the electron energy states in the indirect conduction valleys (L valleys) [15]. Many efforts have been made on that. Delta-doped layer and gas immersion laser doping were used to achieve up to $5\times10^{19}$ cm$^{-3}$ activation of P doping [18-20]. The spin-on dopant process and multiple implantation were also successful in doping Ge up to $1\times10^{20}$ cm$^{-3}$ [22-23]. Laser thermal annealing has been successfully employed to achieve a doping activation above $10^{20}$ cm$^{-3}$ in implanted bulk Ge [24] and in-situ doped Ge-on-Si epilayers [25]. For doping from external sources such as the dopant diffusion from delta-doping method, a high temperature annealing step is commonly used to drive dopants in and activate them. Pellegrini et al. achieved an activated carrier concentration ranging from $2.5\times10^{19}$ cm$^{-3}$ to $2.1\times10^{20}$ cm$^{-3}$ by incorporating P dopants during the Ge growth followed by annealing [26]. For in-situ doped Ge, defect annealing such as thermal cycling of high and low temperature (HT/LT) is commonly used to reduce threading dislocation density (TDD) right after the Ge film epitaxy step. The high thermal budget associated with these steps is not desired, as it drives Si-Ge interdiffusion as well as dopant out-diffusion [27], and thus counteracts the efforts in bandgap engineering. Especially, interdiffusion is enhanced with n-type doping by a factor of 2 to 6 with P doping and 1.5 to 3 times with As doping in low to mid-$10^{19}$ cm$^{-3}$ range [28-30].

Lee et al. studied the impact of high concentration arsenic (As) on Ge epitaxial film grown on Si (001) with 6° off-cut. The observation was that the TDD was reduced by at least one order of magnitude to $< 5 \times 10^6$/cm$^2$, which was attributed to the enhancement in the velocity of the dislocation motion in an As-doped Ge film [3]. However, Ge films without HT/LT annealing were not studied. Recently, Zhou et al. systematically studied the doping and defect annealing impact on Ge film quality for P and As doped Ge on Si (001) with 6º off-cut. The finding was that P and As doping can reduce etch pit density (EPD) without the HT/LT defect annealing step. This suggests that n-type doped Ge films without defect annealing have lower defect density and minimized Si-Ge interdiffusion, which are very promising for Ge light emission. These n-doped Ge films also have smooth morphology and tensile strain of about 0.16 to 0.2%. However, photoluminescence (PL) measurements of the n-doped Ge were not tested, and Ge grown on on-axis Si (100) wafers were not studied either.

This work is a follow-up and more in-depth study of Zhou et al.'s recent work [28]. 1) This work provides photoluminescence (PL) measurements and analysis of n-Ge films on (100) Si substrates without HT/LT annealing, which are more relevant to light emission and to the mainstream Si wafer type, and 2) Zhou et al.'s work used EPD to characterize TDD, which was effective for undoped Ge, but was not suitable for n-Ge [32]. In this work, electron channeling contrast imaging (ECCI) was used to measure TDD of n-Ge [33-34] to provide much more accurate TDD measurements.

# EXPERIMENT DESIGN, RESULTS AND DISCUSSIONS

## Structure Design, Growth and Defect Annealing.

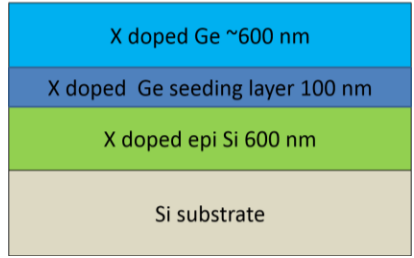

**Figure 1** Schematic diagrams of the 7 samples grown by MOCVD. X stands for As or P.

**Table 1**. Samples without defect annealing were noted as "NA" in the names. "5TC" means with five thermal cycling for defect annealing. "MHT" means the defect annealing was performed only with the merged high temperature steps. "off" means on 6° off-cut (100) Si, and "on" means on-axis (100) Si.

|  | Wafer offcut | Growth method | Dopant and doping technique | Doping level in Ge (cm$^{-3}$) |
|---|---|---|---|---|
| **P-NA-off** | 6º | MOCVD | P, in-situ | $7\times10^{18}$ |
| **P-NA-on** | 0º | MOCVD | P, in-situ | $7\times10^{18}$ |
| **P-5TC-off** | 6º | MOCVD | P, in-situ | $2\times10^{19}$ |
| **A-NA-off** | 6º | MOCVD | As, in-situ | $1\times10^{19}$ |
| **A-NA-on** | 0º | MOCVD | As, in-situ | $1\times10^{19}$ |
| **A-5TC-off** | 6º | MOCVD | As, in-situ | $2\times10^{19}$ |
| **A-MHT-off** | 6º | MOCVD | As, in-situ | $2\times10^{19}$ |
| **MIT Ge control** | 0º | UHVCVD | P, delta doping with a drive-in annealing step | $4\times10^{19}$ |

Seven n-doped Ge-on-Si samples with 3 different annealing conditions (no annealing, 5 HT/LT thermal cycles, and merged HT annealing) were investigated (**Table 1**). The schematic structure is shown in **Fig. 1.** The sample matrix was designed to study the impact of the defect annealing and the 6 degree offcut. There were more samples without annealing, as they were shown to have higher PL than the annealed ones. The merged HT annealing was shown to be equivalent to the 5 HT/LT annealing in our previous study [28].The "MIT Ge control" sample was from the first electrically pump Ge laser work published in 2012 [16] to serve as a benchmark.

All 7 samples were grown on 8-inches Czochrolski (CZ) Si wafers in a metal-organic chemical vapor deposition (MOCVD) tool and the model is CRIUS CCS from Aixtron. The Si substrates are either on-axis (100) Si or (100) Si with 6° off-cut towards the [110] direction. P and As doping levels were chosen as the highest concentration achievable in the epitaxial growth tool. The "MIT Ge control" sample was grown by an ultra-high vacuum CVD tool (UHVCVD) at MIT. The CVD tool model is Sirius 300 from Unaxis. The doping of the "MIT control sample" was conducted with a delta doping technology, where multiple atomically thin P layer were deposited on top of Ge and was later driven in with a high temperature annealing step [16].

For the 7 samples grown by MOCVD, before a Si layer was deposited, the Si substrate was treated at 1050 ± 10 °C for 10 minutes under $H_2$ ambient at 400 mbar. Then, a 600 nm Si layer was grown at 950 ± 10 °C under $H_2$ ambient at 100 mbar. To improve the Ge film quality and reduce the threading dislocations caused by the Ge-Si lattice difference, a 100 nm Ge seeding layer was grown at 400 ± 10 °C under $H_2$ ambient at 100 mbar (low-temperature Ge growth) on top of the Si layer. Finally, a Ge film about 600 nm was grown at 650 ± 10 °C under $H_2$ ambient at 100 mbar (high-temperature Ge growth). We denote these layers as "the top Ge layers" in the discussion below to differentiate from the Ge seeding layers. Immediately after the growth procedure, some of the samples were annealed inside the growth tool while another half were left unannealed (NA) for comparison. Post-deposition thermal cycling was performed by repeating a $H_2$ annealing cycle between low temperature (LT) and high temperature (HT) ranging from 600 °C to 850 °C for 5 times (5TC). Each annealing step was 10 minutes at HT and 5 minutes at LT to improve the quality of the Ge epitaxial film. The ramping rates for heating and cooling were around

1°C/s. Besides, for Sample A, we performed the merged high temperature (MHT) anenaling, namely, with 850°C for 50 minutes anneal with no LT steps to check the difference between HT/LT thermal cycling and merged HT annealing. The temperatures quoted above were the nominal setting temperatures of the MOCVD reactor. After we obtained the Secondary Ion Mass Spectrometry (SIMS) data and compared with well-established interdiffusion model by Dong et al. [20], we found that 890 °C is the best fitting temperature using Dong et al.'s model for Sample P-5TC. Thus, we consider that as a calibrated experimental annealing temperature. For the sample A, since the surface temperature measured was about 20 °C lower than others, we consider the calibrated annealing temperature was 870°C for Sample A-5TC-off and A-MHT-off.

**Surface Roughness Characterization.** Atomic Force Microscope (AFM) measurements were performed to obtain the surface roughness. The scanning area size was 5 μm × 5 μm. Several areas of each sample were measured. The roughness of the samples are around 0.3 - 1.5 nm (**Table 2**), which depends on the area selected and calculation methods. Different dopant configurations or annealing procedures have no significant effect on the surface roughness. The smooth surface is suitable for applications in Ge lasers and as Ge transition layers between GaAs and Si.

**Table 2:** Wafer offcut information and the average and root mean square (RMS) surface roughness of the samples

|  | Wafer offcut | RMS roughness (nm) |
|---|---|---|
| **P-NA-on** | 0 | 0.43 ± 0.13 |
| **P-NA-off** | 6° | 0.57 ± 0.14 |
| **P-5TC-off** | 6° | 1.49 ± 0.48 |
| **A-NA-off** | 6° | 0.32 ± 0.03 |
| **A-NA-on** | 0 | 1.39 ± 0.37 |
| **A-5TC-off** | 6° | 0.35 ± 0.01 |
| **A-MHT-off** | 6° | 1.43 ± 0.69 |

**Strain Measurements by X-ray Diffraction.** High resolution X-ray diffraction (HRXRD) measurements were performed to measure the Ge strain levels using a PANalytical X'Pert PRO MRD with a triple axis configuration. Strain values of the Ge layers are extracted by fitting with the PANalytical Epitaxy software package. The accuracy of this method is ±0.05% for strain measurement in Ge.

The (0 0 4) Ω–2θ scan results are shown in **Fig. 2.** The position of the Ge peaks of all samples are under biaxial tensile strain in comparison with fully relaxed Ge peaks. For the unannealed samples, the associated degrees of relaxation R calculated by combining Eq. (3) and (4) from Ref. [35] are within the range 104% to 105%. This means that the Ge layers are in a slightly tensile strained configuration (~ 0.2%), which is in agreement with literature results. [35-37] The tensile strain is thermally induced to the Ge epilayer during cooling from high-temperature growth or thermal annealing steps to room temperature. In the temperature range of 20 °C to 650 °C, Ge has a coefficient of thermal expansion (CTE) of 5.8 – 8.1 ppm/°C larger than that Si, which is 2.6 – 4.1 ppm/°C [38].

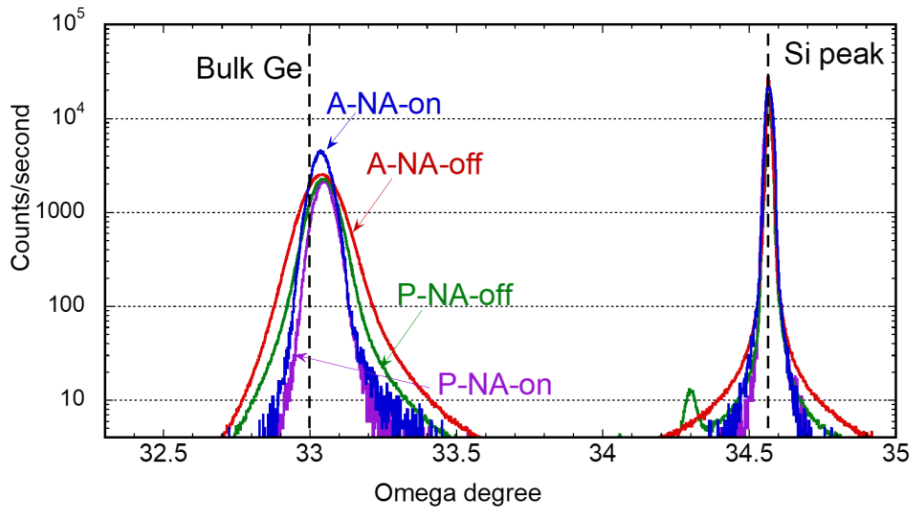

**Figure 2**. HRXRD results of the samples without annealing. The results show that the Ge layers are almost fully strained relaxed. For Ge-on-offcut-Si samples, to exclude the offcut impact on the XRD peaks, the incident X-ray directions were chosen to be perpendicular to the plane formed by the wafer surface normal and the (001) plane normal vectors.

**Threading Dislocation Density Characterization.** EPD has been widely used to characterize TDD of undoped Ge [32, 39, 40]. However, for doped Ge, the etch behaviors are quite different. The etch recipes that work for undoped Ge may not work for doped Ge, and the etch rate is also highly dependent on the doping level [32]. ECCI has been proven to be a reliable, fast and non-destructive method for TDD characterizations [33-34]. The stress fields associated with threading dislocations cause a local deformation of the crystal lattice, which can be imaged using ECCI. TDD measured by ECCI for undoped Ge were in good agreement with EPD and transimisson electron microscopy (TEM) results [34]. Therefore, we believe that ECCI can overcome the weakness of EPD methods for doped Ge, and provide more accurate TDD measurements.

In this work, EPD measurements using a typical etch recipe for undoped Ge were performed for comparison with ECCI. Each sample was etched with an iodine (I2) solution. The I2 solution is a mixture of CH3COOH (100ml), HNO3 (40ml), HF (10ml), I2 (30mg) [39]. Optical microscope and scanning electron microscope (SEM) imaging were used to observe and count the etch pits. The etch rate was approximately 40-80 nm/s depending on the dopants and doping level. After etching roughly half of the top Ge layer (about 300 nm), 4 to 6 different positions on the surface were imaged to calculate an average EPD value. For As and P doped 600 nm-thick Ge films without annealing, EPD values are in the 105 to 106 cm-2 range, which are much lower than expected TDD due to pit merging and the dependencesensitivity on the etch recipe and doping.

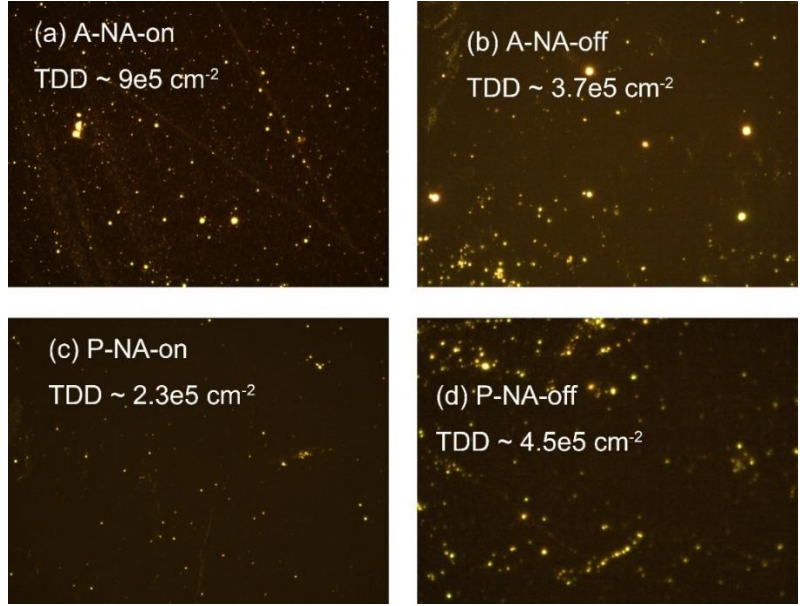

**Figure 3** Optical images for EPD measurements. (a) Sample A-NA-on with 15 s etching; (b) Sample A-NA-off with 15 s etching; (c) Sample P-NA-on with 15s etching; and (d) Sample P-NA-off with 15s etching.

**Table 2.** Summary of the EPD values

| Sample | EPD results (cm$^{-2}$) | ECCI results (cm$^{-2}$) |
|---|---|---|
| **P-NA-on** | $2.7 \pm 1 \times 10^5$ | $4 \pm 0.8 \times 10^8$ |
| **P-NA-off** | $3 \pm 1.5 \times 10^5$ | $5.2 \pm 1 \times 10^8$ |
| **P-5TC-off** | $1.75 \pm 1 \times 10^5$ | $1.1 \pm 0.2 \times 10^8$ |
| **As-NA-on** | $7.3 \pm 5 \times 10^5$ | $2.8 \pm 0.6 \times 10^8$ |
| **As-NA-off** | -- | $1.1 \pm 0.2 \times 10^9$ |
| **As-5TC-off** | -- | $1.5 \pm 0.3 \times 10^8$ |

The ECCI measurements were performed using equipment and conditions similar to Ref. [31]. We used a Thermo Scientific Apreo SEM from Thermo Fisher equipped with a retractable solid state concentric backscatter (CBS) detector inserted below the pole piece of the electron column. For optimum spatial resolution and signal-to-noise ratio, beam energies between (5–20) keV and beam currents ranging from (0.4–3.2) nA were used. A magnetic immersion lens has been used to allow for maximum spatial resolution. The selected diffraction vector was [2 2 0]. The diffraction vector was [220]. The imaging areas were at least 40 square microns. ECCI micrographs are shown in **Fig. 4.**

**Figure 4**.ECCI micrographs (to be added).

**Table 2** compares the EPD and ECCI results on TDD characaterizations of the n-Ge films. It can be seen that the n-Ge films have TDD values above $1 \times 10^8$ cm$^{-2}$, and are commonly three orders of magnitude larger than the TDD by EPD. Defect annealing helped to reduce TDD P-doped Ge.

**Ge and Dopant Concentration Profiling.** Ge and dopant concentrations have a direct impact on the band structure and light emission properties, and are needed in the discussions of PL results. SIMS measurements were performed by Evans Analytical Group to obtain the Ge and dopant profiles. The samples were sputtered with 1 KeV Cs$^+$ primary ion beam obliquely incident on the samples at 60° off the sample surface normal. The sputter rate was calibrated using stylus profilometer measurements. The measurement uncertainty in $x_{Ge}$ is ±1%.

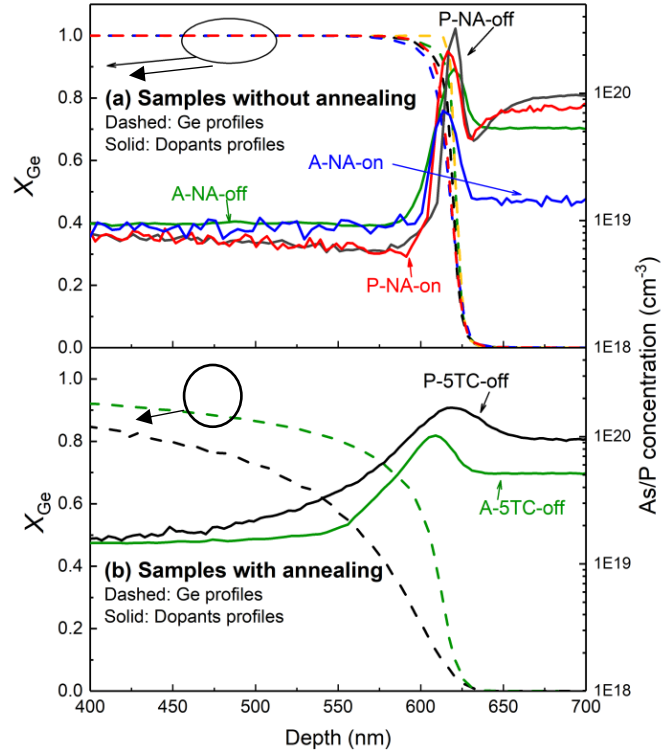

**Figure 4.** Ge molar fraction and dopants (As/P) concentration profiles measured by SIMS. The Ge profiles are shifted laterally for easy comparison. The dopant profiles are also shifted laterally by the same length as their corresponding Ge profiles. (a) n-Ge/Si without annealing, and (b) selected n-Ge/offcut-Si after annealing.

As seen in **Fig. 4 (a)**, the interdiffusion is minimal for the unannealed samples. The 6 degree offcut has little impact on the Ge and dopant profiles. Some differences can be seen in the concentration of the P and As segregation peaks concentrations at Ge/Si interfaces, which may be due to SIMS uncertainty at the interfaces. The SIMS data of annealed n-Ge/offcut-Si samples are in **Fig. 4 (b)**, where significant interdiffusion can be seen. According to the diffusion theory, diffusion and thus interdiffusion is isotropic for cubic crystals. Therefore, we expect that the Ge profiles of n-Ge/on-axis-Si to be the same as those of Ge/offcut-Si. Due to the interdiffusion, the sharp Ge/Si interfaces changed to thick alloy regions. In the full $x_{Ge}$ range, the effective interdiffusivity of Sample P ($\widetilde{D}_P$) is 1.5 to 3 times higher than that of sample A ($\widetilde{D}_A$), and $\widetilde{D}_A$ is 1.5 to 2 times higher than that of Sample U ($\widetilde{D}_U$) [28].

**PL Characterization.** The samples were characterized for PL using a Horiba LabRAM HR Evolution instrument with a Symphony II InGaAs detector. A 1064 nm laser was used on all the samples. The excitation power is ~20 mW at the surface of the sample, and a 50X near infrared objective lens was used. The acquisition time was 6 seconds per spot. The grating used was 300 gr/mm. In order to reduce the noise, first a measurement using "signal" mode was taken, followed by a measurement with the same conditions and under dark condition. The data was then obtained using "signal minus dark" mode. This process was repeated for the measurement of each sample.

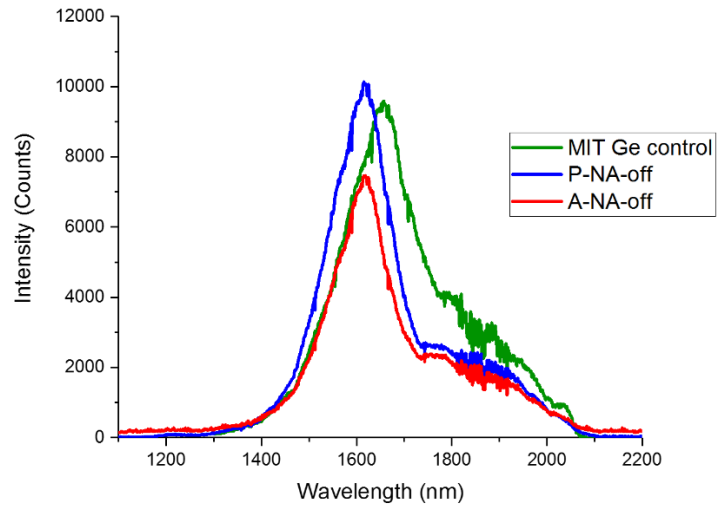

**Figure 5**. PL Measurements comparison for the unnanealed P and As samples and the reference n+Ge control sample from MIT.

The PL characterization results are summarized on **Fig. 5** above. Note that the As and P doped samples are blue shifted compared to the MIT Ge control. This is due the lower dopant concentration of the samples ($1\times10^{19}$ and $7\times10^{18}$ for the As and P doped samples respectively and $4\times10^{19}$ for the MIT Ge control sample as shown on **Table 1**). Higher doping leads to band narrowing[41], therefore the redshift in PL peak positions. On the other hand, the PL intensities of the P and As doped samples are similar to that of the MIT samples despite of the lower n-doping concentration. This result suggests better material quality from the MOCVD grown samples compared to the delta-doped sample. The slight difference in PL intensity between the As and P doped sample could be attributed to the higher EPD value of the As-doped unannealed Ge. **Fig. 6** below shows a comparison for the PL of the annealed and un-annealed samples.

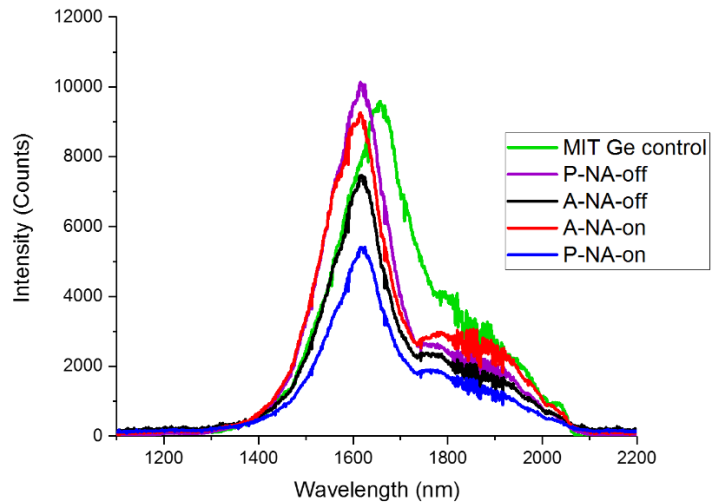

**Figure 6**.PL Intensity for annealed and unnanealed samples for P and As doped samples in comparison with the MIT n-Ge-sample.

The blue shift observed on the annealed samples is mostly due to the Si-Ge interdiffusion, given that the dopant concentration did not decrease after annealing as shown on **Fig. 4**. This is also evidenced by the fact that the indirect transition is enhanced relative to the direct after the annealing, which is observed by a decrease on the relative intensity of direct transitions on the annealed sample. The reduced overall intensity of the annealed sample can be explained by the fact that Si diffused to Ge makes the material more indirect. Regarding the intensity of the on-axis and off-axis samples, this is related to the differences in the dislocation density between the samples. For the P doped sample, the dislocation density seems to be lower for the on axis sample and for the As doped sample the dislocation density for the on-axis sample is slightly higher on average, as shown in **Table 2**. However this has little impact in comparison to the annealing process and Si-Ge interdifussion**. Fig. 7** below shows the impact of the annealing process and TDD as measured by ECCI on the PL intensity.

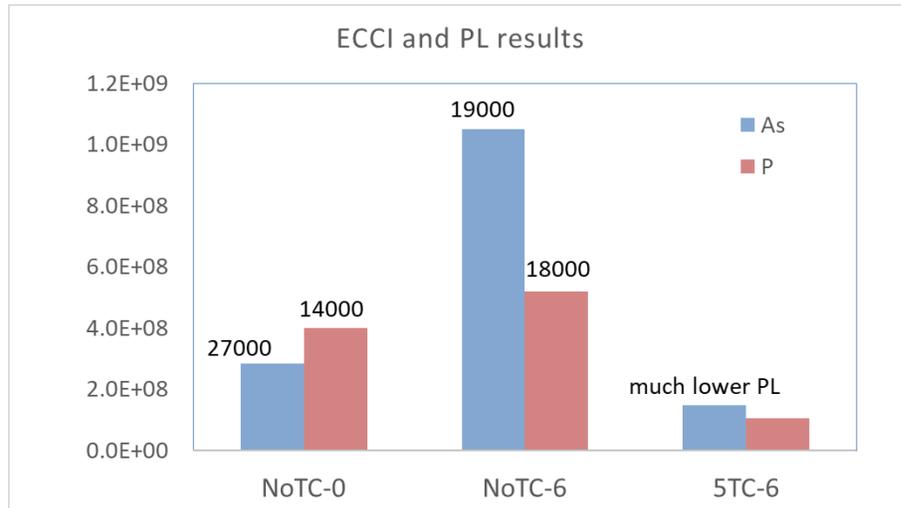

**Figure 7**. TDD values by ECCI and the peak PL intensity comparison. The values above the bars are the peak PL intensity.

Table 3 below shows the energy difference between the direct and indirect band gaps for different samples. The values were obtained by doing a Lorentzian fit on the peaks of the PL spectra and obtaining peak positions using the Origin software. An example of the fitting is shown on **Fig. 7** below. The energy difference between the direct and indirect gaps as well as the (red) shift on emission (band gap narrowing) shows the differences on dopant concentration between the different samples.

**Table 3.** Energy difference between the direct and indirect bandgap for the different samples

| Sample | Direct peak position (nm) | Indirect peak position (nm) | Direct peak position (eV) | Indirect peak position (eV) | Energy Difference (eV) |
|---|---|---|---|---|---|
| A-NA-off | 1610.3 +/- 0.12 | 1854.4 +/- 0.86 | 0.7700 | 0.6687 | 0.1013 |
| N+Ge | 1643.5 +/- 0.26 | 1761.9 +/- 1.5 | 0.7545 | 0.7038 | 0.0507 |
| P-NA-on | 1612 +/- 0.12 | 1841.7 +/- 0.81 | 0.7693 | 0.6733 | 0.0960 |
| P-NA-off | 1609.2 +/- 0.11 | 1880.6 +/- 0.81 | 0.7706 | 0.6594 | 0.1112 |
| P-5TC-off | 1512.6 +/- 0.27 | 1779.6 +/- 0.47 | 0.8198 | 0.6968 | 0.1230 |

Given that there is a linear dependence between the Γ-point bandgap shrinkage and heavy n-doping [29], the band narrowing can be predicted using the dopant concentration given on **Table 1** for each of the samples. Using a first order phenomenological model for band gap narrowing and using the values of $E_{BGN} = 0.013 eV$ and $\Delta_{BGN} = 10^{-21} eV/cm^{-3}$ given for Ge [41], the narrowing was calculated and is summarized on **Table 4** below. Furthermore, it is possible to calculate the dopant concentration from the observed energy difference between the direct and indirect bandgaps on **Table 3**. The results are also shown on **Table 4** as follows.

**Table 4.** Calculated dopant concentration and BGN

| Sample | Measured dopant Concentration | Calculated BGN (eV) | Measured Energy difference (eV) | Calculated Dopant concentration (cm$^{-3}$) |
|---|---|---|---|---|
| P-NA-off | 7×10$^{18}$ | 0.083 | 0.1112 | 9.8×10$^{18}$ |
| P-NA-on | 7×10$^{18}$ | 0.083 | 0.0959 | 8.3×10$^{18}$ |
| P-5TC-off | 2×10$^{19}$ | 0.213 | 0.1230 | 1.1×10$^{19}$ |
| A-NA-off | 1×10$^{19}$ | 0.113 | 0.1014 | 8.8×10$^{18}$ |
| A-NA-on | 1×10$^{19}$ | 0.113 | 0.1114 | 9.8×10$^{19}$ |
| A-5TC-off | 2×10$^{19}$ | 0.213 | 0.1124 | 9.9×10$^{19}$ |
| MIT Ge control | 4×10$^{19}$ | 0.413 | 0.0507 | 3.8×10$^{18}$ |

Note that the calculated band gap narrowing shows agreement with the measurements above and that the calculated dopant concentration agrees with the measured concentrations within an order of magnitude.

**Summary.** To summarize, this work studied thin film quality and photoluminescence of n-doped epitaxial Ge-on-Si. Surface roughness, strain, etch pit desnity, Si-Ge interdiffusion, dopant diffusion, and photoluminescence were characterized to study the impacts of defect annealing and Si substrate offcut on the Ge film quality and most importantly, the light emission properties. All samples have a smooth surface (roughness < 1.5 nm), and the Ge films have a small tensile strain. Six degree offcut was shown to have little impact. Defect annelaing was shown to decrease the photoluminescence intensity greatly due to Si-Ge interdiffusion, which also causes a blue-shift in the emission wavelength. Compared to a benckmarking sample from the first Ge laser work in 2012, the P or As-doped Ge films have similar photoluminescence intensity, smoother surface and less doping concentrations, which are promising for Ge lasers with better light emission efficiencies.

**ACKNOWLEDGEMENTS**

This work was funded by the Natural Science and Engineering Research Council of Canada (NSERC), the National Science Foundation under the Grant No. DMR-1255066 in US, and the National Research Foundation Singapore through the Singapore MIT Alliance for Research and Technology's Low Energy Electronic Systems (LEES) IRG, Competitive Research Program (NRF–CRP19–2017–01) and Innovation Grant from SMART Innovation Centre.

Mr. Libor Strakos, Mr. Ondrej Machek, Drs. Tomas Vystavel and Richard Young from Thermo Fisher, Drs. Andreas Schulze, Clement Porret, Roger Loo at IMEC and Mr. Jacob Kabel at the University of British Columbia are acknowledged for their help in ECCI measurements. Dr. Mario Beaudoin from the Advanced Nanofabrication Facility at the University of British Columbia is acknowledged for the training in HRXRD and EPD measurements and helpful discussions.